| | |
|---|---|
| Title of the paper: | **GMACO-P: GPU assisted Preemptive MACO algorithm for enabling Smart Transportation** |
| Author's name: | Vinita Jindal[1,*], Punam Bedi[2] |
| Institutional Affiliation: | [1,2]Department of Computer Science, University of Delhi, Delhi, India |
| Email Address: | [1]vjindal@keshav.du.ac.in, [2]punambedi@ieee.org |
| Corresponding Author: | [*]Vinita Jindal, |
| Email-Corresponding Author: | vjindal@keshav.du.ac.in |
| ORCID ID: | https://orcid.org/0000-0002-0481-4840 |
| Researcher ID: | A-8520-2016 |
| Scopus Author ID: | 56110234600 |

**Abstract**

Vehicular Ad-hoc NETworks (VANETs) are developing at a very fast pace to enable smart transportation in urban cities, by designing some mechanisms for decreasing travel time for commuters by reducing congestion. Inefficient Traffic signals and routing mechanisms are the major factors that contribute to the increase of road congestion. For smoother traffic movement and reducing congestion on the roads, the waiting time at intersections must be reduced and an optimal path should be chosen simultaneously. In this paper, A GPU assisted Preemptive MACO (GMACO-P) algorithm has been proposed to minimize the total travel time of the commuters. GMACO-P is an improvement of MACO-P algorithm that uses the harnessing the power of the GPU to provide faster computations for further minimizing the travel time. The MACO-P algorithm is based on an existing MACO algorithm that avoid the path with the congestion. The MACO-P algorithm reduces the average queue length at intersections by incorporating preemption that ensures less waiting time. In this paper, GMACO-P algorithm is proposed harnessing the power of GPU to improve MACO-P to further reduce the travel time. The GMACO-P algorithm is executed with CUDA toolkit 7.5 using C language and the obtained results were compared with existing Dijkstra, ACO, MACO, MACO-P, parallel implementation of the Dijkstra, ACO and MACO algorithms. Obtained results show the significant reduction in the travel time after using the proposed GMACO-P algorithm.

## 1. Introduction

NVIDIA Inc. invented the Graphic Processing Unit (GPU) in 1999 to accelerate the computing speed of CPU operations (Sanders and Kandrot 2010). GPU contains several independent processing units with multiple cores or Compute Unified Device Architecture (CUDA) units used for general purpose parallel computation. The GPU has various independent cores also known as Compute Unified Device Architecture (CUDA) units that are accountable for general purpose parallel computations. Using CUDA, we can develop parallel algorithms to be implemented using GPUs. Parallel algorithms speed up the search process by using several computing elements along with a new exploration pattern useful to improve the resulting quality of the sequential algorithms (Pedemonte, Nesmachnow and Cancela 2011), (Gong, et al. 2016). Usually, a program written with GPU support is numerous times quicker than that of the program written with the CPU support. VANETs are integrated in intelligent transportation systems (ITS) to enable smart transportation (U.S. Department of Transportation 2004). In VANETs, the large number of vehicles are present with a very dynamic topology and hence necessitate fast processing to quicken the responses. Another area for research in VANETs is to reduce the time spent on the roads during a journey that includes travel time as well as waiting time at the intersections for a commuter.

Travel time can be reduced by using the shortest path between any two nodes, but if all vehicles select the shortest route using a standard Dijkstra algorithm, then it may create congestion on that route (Dijkstra 1959). Thus, the optimal route is not always the shortest one (Bell and McMullen 2004). Commuters may choose a bit larger route, but their prime preference is to avoid the congestion en route with less number of traffic signals that results in less waiting during the move (Khamis and Gomaa 2014). To solve the real time applications like VANETs, Ant colony optimization (ACO) is used because of its ability to solve complex problems efficiently. ACO is a swarm intelligent metaheuristic algorithm motivated by the social communication behavior of real ants including their self-organization; distributed collaboration and adaptation that helps in solving various real time complex problems competently (Dorigo, Caro and Gambardella 1999), (Jabbarpour, et al. 2014). Ants make use of a chemical substance called pheromones to communicate with other ants to find the optimal path between a food source and their nest.

The value of the pheromone can be adjusted according to the given problem in consideration. In the literature, Modified ACO (MACO) has been presented by Jindal et al (Jindal, Dhankani, et al. 2015) that modifies the behavior of pheromone to reduce the congestion on roads. The MACO algorithm faces the problem that it focusses only on optimization of the route. It does not consider the traffic signals which come in between the selected route and may lead to increase in the total journey time. This problem has been handled in preemptive MACO (MACO-P) algorithm given by Jindal et al (Jindal and Bedi 2017). In their work, authors have considered a less number of vehicles and the MACO-P algorithm works fine for them. But, when the number of vehicles increases, this algorithm get slower due to the serial processing on of a single CPU. Hence, in for providing the quickest computations, A GPU assisted preemptive MACO (GMACO-P) algorithm has been proposed in this paper.

GMACO-P algorithm is improving MACO-P algorithm and intended to run with the parallel hardware. It utilizes the power of GPU computing using CUDA for providing quicker results and assists the commuter to react faster that reduces the total travel time. For implementation, C language is used on CUDA toolkit 7.5 with NVIDIA GEFORCE 710M. The proposed algorithm has been tested on a real time map of North-West Delhi, India. Obtained results show the significant reduction in the total travel time under dense traffic conditions.

## 2. Compute Unified Device Architecture (CUDA) overview

For providing direct access to the GPU's parallel computational elements and virtual instruction sets, CUDA platform is used. In 2006, NVIDIA introduces a general purpose parallel computing programming model and architecture known as Compute Unified Device Architecture (CUDA) (Sanders and Kandrot 2010). Kirk et al stated that both general-purpose multi-core CPUs and multi-thread GPUs have diverse design principles (Kirk and Wen-mei 2012). While CPUs are working for a single thread and providing minimum execution latency, GPUs is working with a large number of threads and maximize the total execution throughput. Thus, for proper resource utilization, a code needs to be implemented using both CPU and GPU simultaneously.

CUDA-capable GPU is performing the actual computations with multiple and highly threaded streaming multiprocessors (SMs). Each SM consists of a number of CUDA cores (also known as streaming processors (SPs) and has its own registers, execution pipelines, cache memories and control units that are shared by all CUDA cores. GDDR DRAM (or global memory) is a high bandwidth off-chip memory used for swift computations and different from the system DRAMs. Figure 1 shows the CUDA execution model with respect to both software and hardware. Software model shows the threads that grouped into blocks that are further grouped into grids. These grids are mainly responsible for performing all executions.

CUDA has the capability to access memory in a way that reduces the data requests to and from the global memory and lower the communication required for data transfer as shown in Figure 2. The host can access the global, constant and texture memories through various CUDA API functions. All threads in the system can access these memories during the execution. Global memory tends to have long access latencies and is a dynamic random access memory (DRAM). Constant memory offers smaller latency read-only access in the situation, when all threads want the access of the same location,. Texture memory is also a read-only memory that provides higher bandwidth by reducing memory requests. It is optimized for spatial locality for the applications with memory access patterns. CUDA also provides the registers and shared memory for high-speed access. For keeping frequently accessed variables, individual threads are using the registers by the associated thread. For thread cooperation and sharing, thread blocks are using the shared memory. For developing high-performance parallel applications, one needs to optimize the kernel's memory access model (Sanders and Kandrot 2010). After using the parallel implementations, obtained results show the promise efficient ways to solve any real time complex problem using CUDA (Sanders and Kandrot 2010).

The paper is structured as follows. The literature review is described in the section 2 followed with the proposed GPU assisted MACO-P (GMACO-P) algorithm in section 3. The experimental setup and results are discussed in section 4 followed by the conclusion in section 5.

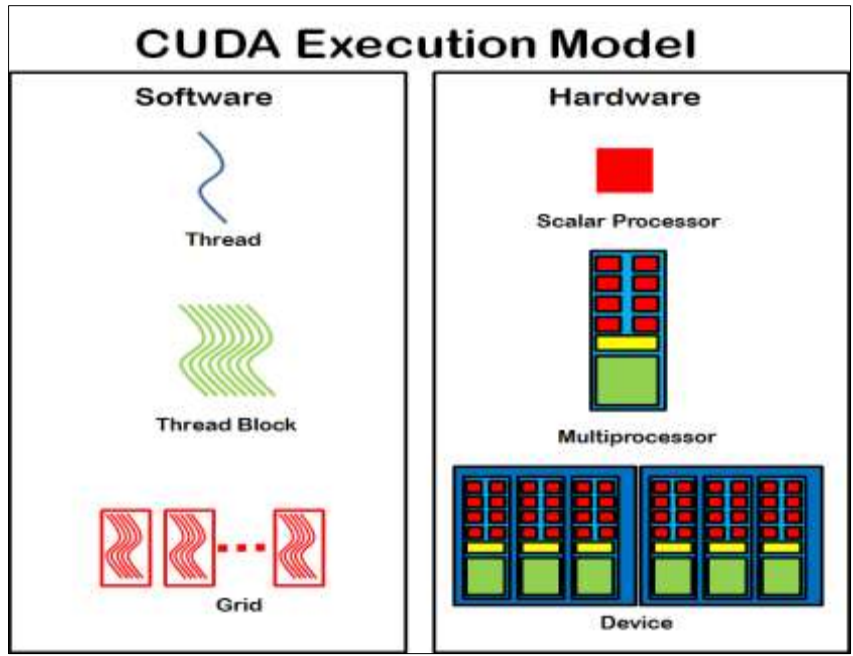

**Figure 1. CUDA Execution Model**

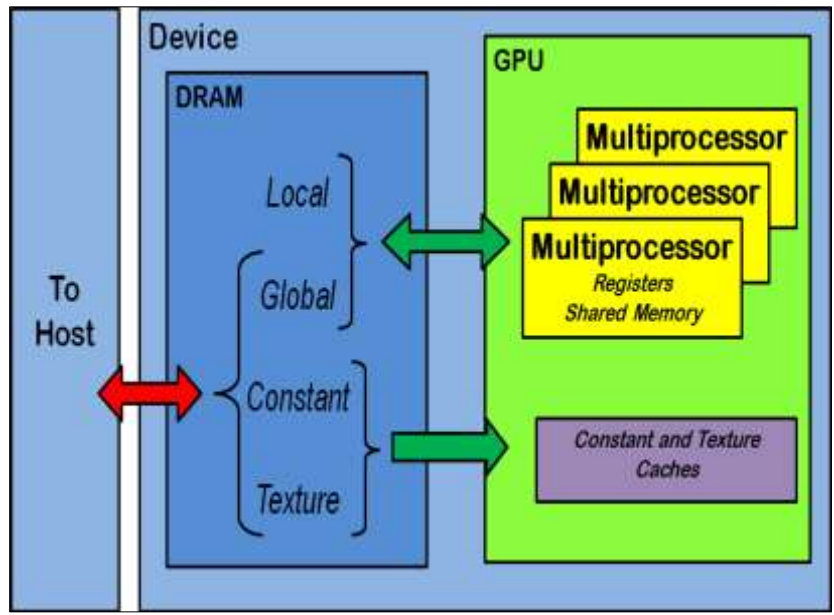

**Figure 2. CUDA Device Memory Model**

## 3. Literature Review

In developing countries like India, commuters spend most of their commuting time either in congestion on roads or waiting for their turn to cross the intersections. This time can be reduced by lowering the congestion en route and/or the waiting time at intersections by optimizing traffic signals and proving the vehicles a congestion free route. Traffic signals are used for managing the traffic at the intersections and can be categorized into fixed signals and adaptive signals (Florin and Olariu 2015). Fixed traffic signals are operated in pre-decided constant timings (Roess, Prassas and McShane 2011). Fixed signals can create the congestion, as one direction may have more traffic in comparison to other directions and these signals do not consider the real time traffic density into consideration in assigning the green time. Adaptive signals can adjust the signal durations according to the real time traffic density on the intersections (Dias, et al. 2014), (Vasirani and Ossowski 2011), (Bedi, Jindal and Garg, et al. 2015). A lot of research in the area of traffic signals is going on to improve traffic signals due to the significant increase in road congestion nowadays (Shivani and Singh 2013), (Falcocchio and Levinson 2015), (Clempner and Poznyak 2015).

Adaptive signals can be used either in preemptive or non-preemptive manner. To the best of our knowledge, most of the adaptive systems are designed in non-preemptive manner in literature. These signals faces the problem of unused green time at the phase with no vehicle resulting in more waiting for a green time assignment for other phases. To alleviate this problem, the preemptive approach can be used for providing out of order green light assignment to the phase with more traffic which can help in reducing the congestion in real time (Bedi, Jindal and Garg, et al. 2015). Their work is handling the real time traffic demand by allotting the green time to the phase with huge number of vehicles out of its fixed cyclic order. Other authors have also implemented the preemptive signals, but their work is considered to handle the emergency vehicles only (Huang, Shiue and Luo 2015), (Wang, et al. 2013). For the hassle free movement of buses at the intersections, and to provide the lesser delays for buses, some of the researchers have used a preemptive approach (Li, Zheng and Li 2016), (Hu, Park and Lee 2016). In this paper, we have used preemptive adaptive signals for reducing the waiting time at the intersections as presented by Bedi et al in their work (Bedi, Jindal and Garg, et al. 2015).

The GPU has been used by researchers to perform quicker computation and providing substantial results. The authors in their work (Wang, Li and Zhang 2015), (Shen, Wang and Zhu 2011), (Laua and Srinivasan 2016) have used the harnessing power of GPU to get faster results in traffic domain. In (Jindal and Bedi 2017), the authors have implemented the preemptive algorithm in parallel on the GPU for traffic control. Their work considered only minimization of waiting time at intersections which works under normal road conditions. To handle the case when there is some problem on the selected path, cannot be handled by their work. Authors in their paper (Jindal and Bedi 2017) have combined the preemptive algorithm with the routing algorithm for the selection of non-congested and optimal route. In this paper, we have combined both traffic and route optimization with their parallel implementation for faster response time.

Routing forms an integral part of the transportation system as it helps commuters to reach the destination faster and in the best possible way. It is a very challenging task due to the variable speed and dynamic mobility patterns of the vehicles (Dias, Rodrigues and Zhou 2014). Various routing algorithms for VANETs have been proposed in literature by a number of authors in their work (Lee, Lee and Gerla 2010), (Chaqfeh, Lakas and Jawhar 2014), (Dua, Kumar and Bawa 2014). Dijkstra's algorithm is the standard routing algorithm that finds the shortest path between any two nodes in a network (Altayeb and Mahgoub 2013), (Al-Sultan, et al. 2014). The algorithm works on local optimum solution such that it does not target the destination from the start rather it tries to find the next best possible node from the current node and marks them visited to avoid the cycles. Dijkstra algorithm works well for less number of vehicles. But, the computation time for the algorithm increases significantly as the number of vehicles increase in the VANETs. To solve the routing problems in such cases, normally swarm intelligence techniques are used (Lee, Lee and Gerla 2010), (Altayeb and Mahgoub 2013), (Bedi, Mediratta, et al. 2007).

Swarm intelligence techniques replicate the behavior of existing living organisms and are also used in the literature for routing in VANETs (Dempsey and Schuster 2005). Swarms have the capability to answer any complex problems and hence are suitable for the real time traffic scenario. Ant colony optimization (ACO) is an artificial swarm algorithm that can be used for building path in communication and networking domains. It has the ability to develop self-organizing methods that help in solving real time routing problems (Dorigo, Caro and Gambardella 1999), (Claes and Holvoet 2011), (Geetha, Vanathi and Poonthalir 2012), (Dorigo and Stützle 2004). In ACO, pheromone left by an ant on its route while passing through that route is sensed by other ants in order to follow the traversed route with maximum pheromone (Cabanes, et al. 2014). Quadratic assignments and travelling salesman problems are some initial problems that used ant colonies optimization technique (Jindal, Dhankani, et al. 2015). After this, several domains also exist that used an ACO technique for vehicle routing problem in the literature (Bell and McMullen 2004). One of the approaches discussed in literature for dynamic adaptation of traffic signals using graphs is presented in (Bedi, Mediratta, et al. 2007). A modified version of ACO for the vehicular traffic environment to yield more optimized results on complex road network is presented in (Bin, Zhen and Baozhen 2009). The core objective of using ACO is to minimize the waiting time during journey resulting in reduction of the total travel (Dias, et al. 2014).

For reducing the congestion on roads, a Modified ACO (MACO) is proposed by the authors in their paper (Jindal, Dhankani, et al. 2015). MACO modifies the pheromone behavior and based on the pheromone deposition on the roads, the vehicles have deviated from their chosen path to escape the congestion en route. The authors in their work have not considered the traffic signals en route that may increase the waiting at intersections subsequently an increase in total travel time. To solve this problem, a preemptive algorithm with the MACO algorithm has been combined in (Jindal and Bedi 2017) that results in the reduction of the overall travel time during congestion. Their algorithm is implemented in a sequential mode. The existing system is having huge numbers of vehicle with dynamic topology, and hence a requirement for the faster computation of the algorithm is desired. In this paper, this has been achieved by using the GPU for the parallel implementation of the proposed algorithm.

Despite the fact that the parallel implementation of ACO through GPUs has been used by various researchers (Fu, Lei and Zhou 2010), (Dawson and Stewart 2013), (Cecilia, et al. 2011) in their work, as per best of our knowledge, none of the researcher has implemented the complete algorithm in parallel. The main phases of the Ant Colony Optimization (ACO) algorithm are pheromone deposition, route planning and pheromone updating. Only the route planning phase has been paralleled by most of the early implementations. In (Jindal and Bedi 2016), (Jindal and Bedi 2017) authors have implemented the MACO algorithm with all phases to be run in parallel that reduces the overall processing time for the algorithm, which further lower the total travel time.

The drawback of the work done by the authors (Jindal and Bedi 2016), (Jindal and Bedi 2017) was that MACO-P only considers routing decisions regardless of the presence of the traffic signals on the selected route that may increase the waiting time for the travelers. To solve this, GPU assisted Preemptive MACO (GMACO-P) algorithm is proposed in this paper that parallelizes both preemptive and MACO algorithms in combination. The proposed GMACO-P algorithm is discussed in the next section.

## 4. Proposed GPU assisted MACO-P (GMACO-P) algorithm

During the trip, a commuter spends his major time in waiting due to either congestion on the roads or due to waiting at traffic signals on intersections. Thus, for decreasing the overall travel time, one must decrease the waiting time on both of these areas concurrently. In (Jindal and Bedi 2017), the authors have presented a Preemptive MACO (MACO-P) algorithm in VANETs for reducing travel time during a trip. The MACO-P algorithm reduces waiting time at the intersections along with path optimization by avoiding the congested route. MACO-P algorithm is an improvement of MACO algorithm used for routing that optimizes the traffic signals by incorporating the preemptive algorithm. The problem associated with MACO-P algorithm is the time taken by the algorithm in the case of a large number of vehicles is very high and processing is done on a machine with a single CPU. To alleviate the problem of slow computations, a GPU assisted MACO-P (GMACO-P) algorithm has been proposed in this paper.
In the proposed algorithm, CUDA enabled GPUs are being used for providing fast communication. After implementing the GMACO-P algorithm, the waiting time is minimized for the commuters' at the intersections with non-congested optimized path resulting in less travel time resulting in a reduction in the cost of travelling to reach destination. The existing MACO-P and proposed GMACO-P algorithms are discussed in the following sections.

### 4.1 Existing MACO-P algorithm

The MACO-P algorithm In (Jindal and Bedi 2017) has been designed by improving the MACO algorithm (Jindal, Dhankani, et al. 2015). Several traffic signals encounter on the path selected by the MACO algorithm. These traffic signals (i.e. Fixed or adaptive signals) add the waiting time in total travel time. For reducing the waiting time at the signals, MACO-P is used that incorporates the preemptive traffic signal algorithm (Bedi, Jindal and Garg, et al. 2015) into the MACO algorithm that minimizes the total journey time by minimizing the overall waiting at the signal. In the MACO-P algorithm, first a non-congested path with minimum pheromone is selected by the MACO algorithm and then on that path whenever a traffic signal is encountered, the preemptive traffic signal algorithm has been applied to reduce the waiting time at the intersection
.

```
Pseudocode: Preemptive MACO (MACO-P) (L:Light, P:Phase)

G(N,E) = graph, G with node set, N and edge set, E
Ph_{r(a,b)} = value of pheromone on road, r from node a to node b
N_t = number of vehicles in simulation at time, t
Q_i = Queue Length at each lane of the road for phase, P_i
t_j = Waiting time of vehicle at phase, P_j
for i = 1(1) n
        veh_i.present = getnodeposition(veh_i,G)
        if veh_i.present = veh_i.destin
                i++
        else if veh_i.present is in between the nodes
                node = next_node(veh_i, G)
                road = E (last_node(veh_i, G), next_node(veh_i, G))
        else
                calculate A(veh_i.present) //adjency matrix for veh_i.present ε N
                select node with min (E(veh_i.present, x)) s.t. x ε A(veh_i.present)
                road = E(veh_i.present, node)
                if N_t > threshold
                    select node1 with min (E(veh_i.present, x)) s.t. x ≠ node and x ε A(veh_i.present)
                    road = E(veh_i.present, node1)
                    node = node1
        end if
            if road.L == true //road has traffic signals
              for u = 1 (1) 8
                if Q_u == 0
                    for all values of v s.t. v ≠ u
                                        if all Q_v ≠ 0
                        compute max(Q_v)
                        L_v = 1 for max(Q_v) //Greenlight Assignment for phase, v
                       for all values of w s.t. w ≠ v
                            go to Label
                    end for
                else
                    L_u = 1     //Green light Assignment for phase, u
                    for all values of w s.t. w ≠ v
                        go to Label
                end if
                Label:
                    for any Q_w
                       if Q_w > Th_max
                            L_w = 1           //Green light Assignment for phase, w
                       else if t_w > T_max
                            L_w = 1           //Green light Assignment for phase, w
                       else
                            compute max(Q_w)
                        L_w = 1 for max(Q_w)    //Green light Assignment for phase, w
                    end for
                continue for loop
            veh_i.present = node
            Ph_{road}++ //Increment pheromone at the selected road
            Ph_{[N-road]}-- //Decrement pheromone at all other roads accept the selected road
continue for loop
```

**Figure 3: pseudocode for MACO-P algorithm (Jindal and Bedi 2017)**

The algorithm for the MACO-P algorithm is shown in Figure 3 and is working of the same is given as follows.
- If the vehicle is present in between the two nodes, i.e. on the edge, then the next node on the same path is selected as the next node.
- Otherwise the next node with the minimum pheromone value among all the paths leading to the destination, from the current position to that node between the adjacent nodes is found.
- If the chosen path with the minimum pheromone value has traffic signals, for handling the traffic signals, preemptive algorithm is used.
- After handling the traffic signals and finding the next node, the position of the vehicles is updated.

## 4.2 Proposed GMACO-P algorithm

The MACO-P algorithm was running on a system with a single CPU and hence requires more computation time due to the serial computations for large number of vehicles. To handle the dynamic traffic of VANETs high computation speed, GPU assisted MACO-P (GMACO-P) has been proposed in this paper. It is implemented on CUDA enabled NVIDIA GPUs using C language with Visual studio 2013. The proposed algorithm works by taking the current vehicle density taken from the sensors mounted on the roads for the actual computations. The GMACO-P algorithm reduces the total travel time by providing faster result and leaving time for their decision as all the subparts of the algorithms are designed in a way to run in parallel using device functions. The GMACO-P algorithm is given in Figure 4 and Figure 5 respectively and summarized as below.

In Figure 4, the CUDA_GMACO-P function is given and it is running on the CPU host. In this function, cudaMalloc function is allocating the CUDA memory to all the system variables of the device needed during the computation. Then, cudaMemcpy function with HostToDevice option copies each of the variables to the device independently. Next, CUDA_GMACO-P_KERNEL, the main kernel function, is called with the parameters: grid size and block size of the device grid.

All the tasks of the kernel function has been fully parallelized in order to get faster processing. The result variables need to copied again by cudaMemcpy function with DeviceToHost option. Finally, earlier allocated CUDA memory is deallocated by the cudaFree function. The CUDA_GMACO-P_Kernel function will be evoked on the GPU device; it has seven device functions named as: Calc_Dist, Adj_Neigh, Eval_Rout, Calc_Dens, Comp_Thres, Updt_Pher and Evap_Pher. These functions are evoked by the kernel to perform the various sub functions in parallel. The kernel's job is to assign the parallel threads to be used by each phase, nodes and traffic signals for the networks. The value of vehicle density and the threshold value of the road in each phase is used for the green light assignment. Initially, each path is allocated with initial random pheromone value.

The device function Calc_Dist is calculating the distance between any source to destination for different routes in the graph. Next, the device function Adj_Neigh function is finding all the adjacent neighbours for the current position of the vehicle. Then the minimum value of the pheromone is being chosen for the calculation of the next position and before selecting that nodes next position, threshold is being taken into consideration. The Eval_Rout device function is evaluated for the congestion based on the value of threshold. Then the device function Calc_Dens is calculating vehicle density on the route between current and next position for assigning green light to avoid the waiting. The phase with the maximum queuing is allotted green light. The default order is preserved, in case of absence of such phase. Comp_Thres function is calculating the actual density of vehicles and finds the phase with maximum queuing, it then compares with the value of the threshold for deciding about the allotment of green light. Here, in order to avoid the congestion, the green light assignment can be done without considering the normal order.

Updt_Pher device function is responsible for the updating of the pheromone and the device function Evap_Pher is called for evaporating the pheromone deposition on unused route. After parallelization of all the subtasks, the computation speed will get increased and results are provided at faster speed. The faster results enable the commuters to get extra time for reacting and choose the appropriate action on the traffic signals. This results in lowering the average Queue length and thus reduces the waiting time for the entire journey. Algorithm 1 runs using the CPU version while Algorithm 2 runs using the GPU version. The experimental setup and results obtained is presentsed in the next section.

```
Algorithm 1: CUDA_GMACO-P (graph, distance, route, pheromone, traffic_light, phases)

1. Allocate device memory using cudaMalloc function.
2. Copy inputs from host to device using cudaMemcpy function.
3. Invoke CUDA_GMACO-P_KERNEL <<<grid size, block size>>> (graph, distance, route,
   pheromone, traffic_light, phases)
4. Copy results back from device to host using cudaMemcpy function.
5. Free CUDA memory using cudaFree function.
```

**Figure 4: CUDA_GMACO-P Algorithm**

```
Algorithm 2: CUDA_GMACO-P_KERNEL (graph, distance, route, pheromone, traffic_light,
   phases)

1. In the beginning, the device function Calc_Dist (graph, distance, route) is called to calculate the
   distance among various routes in the graph from any source to destination.
2. Each thread is assigned for each phase, all nodes and traffic signals in the network.
3. Then pheromone for each edge is initialized with some random value.
4. Next, adjacent neighbours are found by the device function Adj_Neigh (graph, current_node,
   pheromone).
5. By using the minimum value of pheromone and threshold, next node is found.
6. Next, device function Eval_Rout (route, distance) is called that evaluate the route for congestion.
7. Invoke device function Calc_Dens (traffic_light, phases) to calculate density of the vehicles at each
   phase of traffic signal.
8. Invoke device function Comp_Thres (phase, traffic_light) to allot the green time.
9. Then, device function Updt_Pher (graph, route, pheromone) is called for updating the value of
   pheromone.
10. Finally, device function Evap_Pher (route, distance) is called to reduce the pheromone deposition.
```

**Figure 5: CUDA_GMACO-P_KERNEL Algorithm**

## 5. Experimental setup and results

The proposed GMACO-P has been simulated using the CUDA toolkit 7.5 with C programming on Microsoft Visual Studio 2013. In the parallel MACO-P (GMACO-P) algorithm, we have used the North-West Delhi network obtained from the Google map. In the implementation, vehicles can communicate directly or indirectly with other vehicles and/or the fixed infrastructure. The information is received either through the GPS or the sensors installed on the roads. This information is used by the GMACO-P algorithm for the selection of routes and threshold for determining the green light allotment for the traffic signals. The snapshot of the North-West Delhi network used for experimentation is given in Figure 6. The network consists of 52 nodes that are connected by 128 roads consisting of 3 lanes. The proposed GMACO-P algorithm is using the GPU is to reduce the computation time needed for both green light assignment and routing decision. This will finally minimize the waiting time and hence lessen the total travel time for the travelers.

In GMACO-P algorithm, a random and uniform speed range of 50 m/s to 80 m/s was used for the vehicle mobility. For the selection of congestion free path, GMACO-P algorithm optimizes the path length and testing was carried out with varied number of vehicles on CUDA enabled GPUs. The experiment was run 5000 times to ensure the accuracy and the average value for the results was taken. During the trip, whenever a traffic signal comes across the path, the preemptive traffic control algorithm has been applied. Results obtained by the proposed GMACO-P algorithm is compared with that of Dijkstra, ACO, MACO, MACO-P and the parallel implementation of Dijkstra, ACO and MACO algorithms under similar environment. The GMACO-P algorithm provides considerable decrement in the total travel time for the vehicles because of the reduction in computation time.

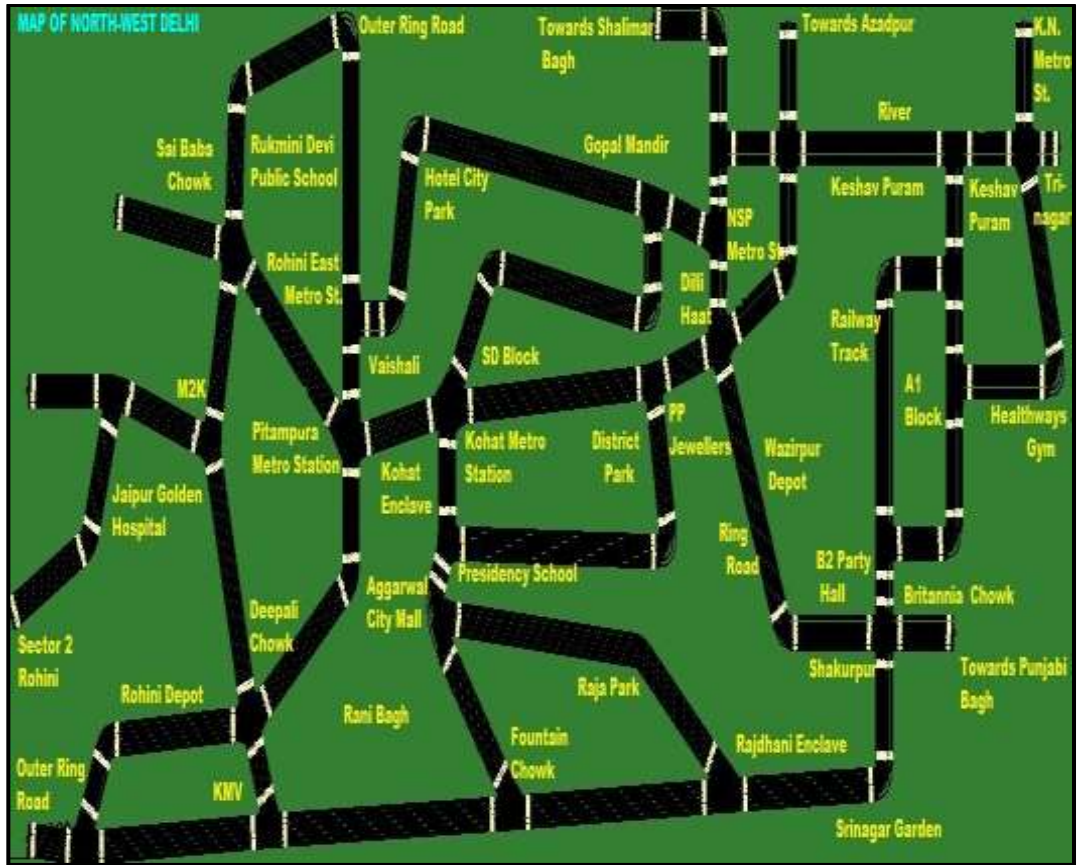

**Figure 6: Screen shot for real time North-West Delhi network**

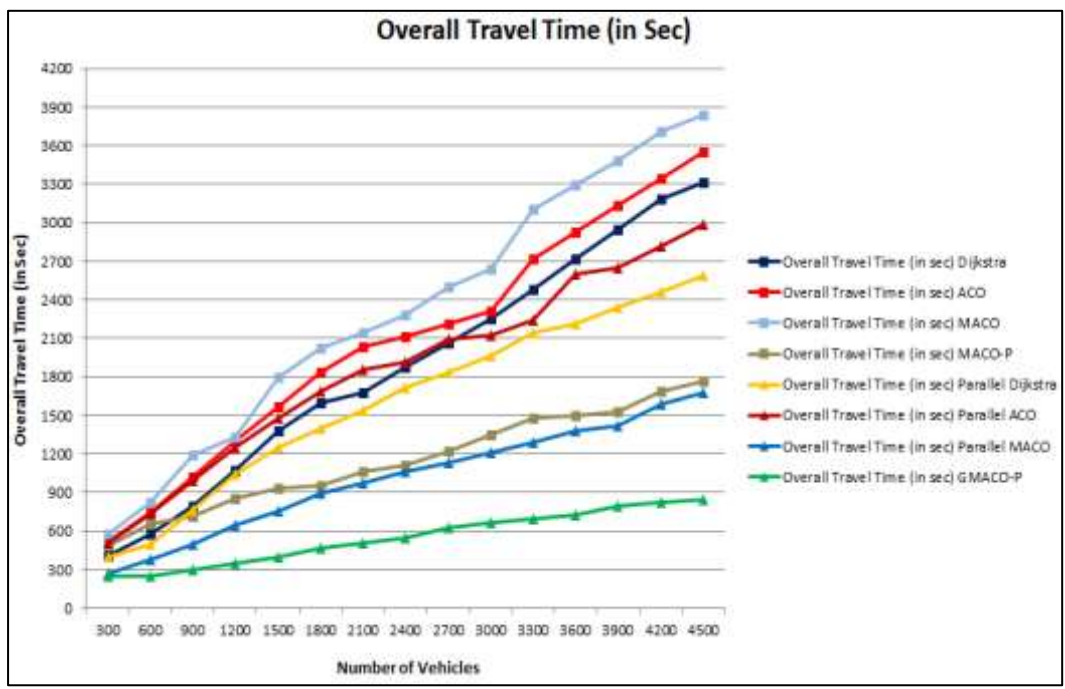

**Figure 7: Graphical representation for Overall travel time in North-West Delhi Network**

Table 1: Results in terms of overall travel time in North-West Delhi Network

| No. of Vehicles | Overall Travel Time (in Sec) | | | | | | | |
|---|---|---|---|---|---|---|---|---|
| | Dijkstra | ACO | MACO | MACO-P | Parallel Dijkstra | Parallel ACO | Parallel MACO | GMACO-P |
| 300 | **413** | **513** | **578** | **487** | **394** | **504** | **273** | **247** |
| 600 | **578** | **742** | **820** | **655** | **502** | **731** | **382** | **253** |
| 900 | **796** | **1025** | **1195** | **719** | **770** | **998** | **498** | **296** |
| 1200 | **1075** | **1303** | **1328** | **855** | **1042** | **1249** | **651** | **345** |
| 1500 | **1382** | **1569** | **1792** | **932** | **1247** | **1484** | **756** | **403** |
| 1800 | **1597** | **1835** | **2026** | **958** | **1395** | **1687** | **895** | **471** |
| 2100 | **1678** | **2034** | **2147** | **1062** | **1536** | **1852** | **973** | **507** |
| 2400 | **1873** | **2117** | **2278** | **1113** | **1718** | **1915** | **1059** | **546** |
| 2700 | **2064** | **2211** | **2498** | **1217** | **1834** | **2093** | **1137** | **624** |
| 3000 | **2256** | **2313** | **2642** | **1349** | **1963** | **2125** | **1209** | **665** |
| 3300 | **2481** | **2718** | **3109** | **1478** | **2147** | **2239** | **1291** | **693** |
| 3600 | **2719** | **2924** | **3295** | **1499** | **2216** | **2597** | **1384** | **728** |
| 3900 | **2950** | **3136** | **3478** | **1524** | **2345** | **2648** | **1416** | **798** |
| 4200 | **3183** | **3342** | **3710** | **1687** | **2458** | **2814** | **1593** | **824** |
| 4500 | **3317** | **3547** | **3841** | **1765** | **2592** | **2986** | **1675** | **845** |

During experimentation, we regularly increased the number of vehicles and noted the total travel time for respective case. The results are shown in Table 1 with their graphical interpretation as shown in Figure 7. It was observed that after using the proposed algorithm with 4500 vehicles; the overall travel time is significantly reduced by 74.53%, 76.18%, 78.00%, 52.12%, 67.40%, 71.70% and 49.55% as compared with Dijkstra, ACO, MACO, MACO-P, parallel implementation of Dijkstra, ACO and MACO algorithms respectively. The obtained results show that GMACO-P algorithm is effectively able to lower the total travel time as compared with its counterparts with increase in vehicles.

## 6. Conclusion

A novel GPU assisted MACO-P (GMACO-P) algorithm is presented in this paper. GMACO-P is an advancement of the existing MACO-P algorithm that runs using NVIDIA GPUs and provide fast computation on multiple parallel processors that enable commuter to respond quickly. GMACO-P algorithm is designed for optimizing both green time assignment at traffic signals and selection of optimal congestion free route for the vehicles on move by using the harness of CUDA enabled GPUs. The GMACO-P algorithm has been executed using the CUDA toolkit 7.5 with C language on the Microsoft Visual studio 2013 with NVIDIA GeForce 710M architecture. For experimentations, the network used is obtained from the Google map of North-west Delhi, INDIA.

The obtained results shows the out performance of the presented GMACO-P algorithm over the existing Dijkstra, ACO, MACO, MACO-P and parallel implementations of Dijkstra, ACO and MACO algorithms. The GMACO-P algorithm shows a significant reduction in the total travel time taken by vehicles as compared with its counterparts with 99% confidence. It was observed that after using the presented algorithm with 4500 vehicles, the overall travel time is significantly reduced by 74.53%, 76.18%, 78.00%, 52.12%, 67.40%, 71.70% and 49.55% as compared with Dijkstra, ACO, MACO, MACO-P, parallel implementation of Dijkstra, ACO and MACO algorithms respectively.